# A high-field $^3$He Metastability Exchange Optical Pumping polarizer operating in a 1.5 T medical scanner for lung MRI

G. Collier,[1,a,b)] T. Pałasz,[1] A. Wojna,[1] B. Głowacz,[1] M. Suchanek,[2] Z. Olejniczak,[3] and T. Dohnalik[1]

[1]*Institute of Physics, Jagiellonian University, Krakow, 30 059, Poland*

[2]*Department of Chemistry and Physics, Agricultural University, Krakow, 31 120, Poland*

[3]*Institute of Nuclear Physics, Polish Academy of Sciences, Krakow, 31 342, Poland*

After being hyperpolarized using the technique of Metastability Exchange Optical Pumping (MEOP), $^3$He can be used as a contrast agent for lung magnetic resonance imaging (MRI). MEOP is usually performed at low magnetic field (~ 1 mT) and low pressure (~ 1 mbar), which results in a low magnetization production rate. A delicate polarization-preserving step of compression is also required. It was demonstrated in sealed cells that high nuclear polarization values can be obtained at higher pressures with MEOP, if performed at high magnetic field (non-standard conditions). In this work the feasibility of building a high-field polarizer that operates within a commercial 1.5 T scanner was evaluated. Preliminary measurements of nuclear polarization with sealed cells filled at different $^3$He gas pressures (1.33 to 267 mbar) were performed. The use of an annular shape for the laser beam increased by 25 % the achievable nuclear polarization equilibrium value ($M_{eq}$) at 32 and 67 mbar as compared to a Gaussian beam shape. $M_{eq}$ values of 66.4 and 31 % were obtained at 32 and 267 mbar respectively and the magnetization production rate was increased by a factor of 10 compared to the best results obtained under standard conditions. To study the reproducibility of the method in a polarizing system, the same experiments were performed with small cells connected to a gas handling system. Despite careful cleaning procedure, the purity of the $^3$He gas could not be matched to that of the sealed cells. Consequently, the polarization build-up times were approximately 3 times longer in the 20-30 mbar range of pressure than those obtained for the 32 mbar sealed cell. However, reasonable $M_{eq}$ values of 40-60 % were achieved in a 90 mL open cell. Based on these findings, a novel compact polarizing system was designed and built. Its typical output is a $^3$He gas flow rate of 15 sccm with a polarization of 33 %. *In-vivo* lung MRI ventilation images (SNR of

---

a) g.j.collier@sheffield.ac.uk

b) Present address: Department of Cardiovascular Science, University of Sheffield, United Kingdom

approximately 55 for a voxel size of 50 mm x 3 mm x 3mm) were acquired to demonstrate the polarizer's application.

## I. Introduction

Hyperpolarized $^3$He is used in a diverse range of applications including neutron spin filters[1-4] and as scattering targets for electrons[5, 6] in nuclear physics, nuclear magnetic resonance[7, 8] and Magnetic Resonance Imaging (MRI) of the lung in animals[9] and humans[10]. In lung MRI, it has been used as a tracer gas with multiple aspects of functional sensitivity available with different MR pulse sequences[11]. It has demonstrated a great potential in the study of a wide range of chest diseases such as asthma[12], cystic fibrosis[13], chronic obstructive pulmonary disease[14], lung cancer[15] and also lung transplantations[16]. Since lung diseases have become a huge cause of mortality in the world (COPD is now the fourth leading cause of chronic morbidity and mortality in the United States[17]), interest in producing faster and larger quantities of hyperpolarized $^3$He for lung MRI has grown rapidly.

Two optical methods for production of hyperpolarized $^3$He exist: Spin Exchange Optical Pumping (SEOP) and Metastability Exchange Optical Pumping (MEOP). The main advantage of SEOP over MEOP is that optical pumping (OP) is performed directly at several bars. However, it is a long process that takes several hours, involves the use of toxic rubidium, and the level of polarization achieved is usually lower than in MEOP. Under standard conditions, MEOP is performed at low pressure (~ 1 mbar) inside a magnetic field of a few mT and polarization values up to 90 % can be obtained in sealed cells[1, 18]. However, achieving a good polarization level and a high production rate in a polarizer is more challenging, due to polarization losses induced by the stage of compression (compression factor of ~1000) required for the main $^3$He applications. Hence, a number of different strategies have been established. In the University of Mainz, large scale and centralized productions have been chosen[1, 19], coupled with a specially designed storage system which provide long relaxation time of $^3$He, allowing the shipment of the polarized gas to different partners[20]. An efficient production rate of 20 to 60 sccm (standard cubic centimeter per minute) has been achieved for a nuclear polarization of 75 and 60 %, respectively. The main drawbacks however are: the price of such a system, due to the cost of the non-magnetic titanium alloy piston compressor; its bulky size (the polarizer containing five OP cells of 2.4 m length for a total volume of 36 L); and the difficulties in accommodating user demands with regards to shipment over large distances. Another approach is to build smaller polarizers, that are easy to handle and storable nearby the MRI scanner or other facilities for on-site production. Some attempts to build a more compact polarizer working with a modified diaphragm pump[21] and an aluminum piston compressor[22] have been successful in the United States. A table-top polarizer has also been designed in Paris using a peristaltic compressor[23] and recently modified by our group[24]. In general, such compact polarizers have the advantage of having lower cost and less constraints but the reduction in size is obtained at the expense of lower gas production rates. They are typically around 3 to 5 sccm in all these systems, for polarization values varying between 30 to 55 %.

The limiting factor for a higher production rate in compact systems is the pressure (~ few mbar) at which MEOP is performed. For a long time, it was thought that MEOP could not produce reasonable polarization values at higher pressures, due to enhanced collisional relaxation processes[25]. However, performing MEOP at 0.1 T was found to have a beneficial effect on achievable nuclear polarization at higher pressures[26]. A factor of 2 increase in nuclear polarization in a sealed cell at 40.4 mbar was reported. Indeed, the influence of hyperfine coupling in the structures of the different excited levels of helium in the plasma discharge is strongly reduced at high magnetic field. Hence, the loss of nuclear polarization due to the transfer of nuclear orientation to electronic spin and orbital orientations is expected to be lower. The first attempts to explain this important improvement[25], plus a new theoretical framework and experimental methods[27], and additional results obtained at 1.5 T[28, 29] were published shortly after. However, the lack of conclusive agreement between experimental results and theoretical values led to a joint collaboration between Kastler Brossel Laboratory and our group, in order to conduct further investigations. Thus, systematic studies of the MEOP process were performed at a wide range of magnetic fields (0.45, 0.9, 1.5, 2 and 4.7 T), pressures (from 1.33 to 267 mbar), densities of the $^3$He metastable state, but also OP transitions, pump laser bandwidths, intensities and beam shapes[30-32]. The fundamentals of MEOP under standard and non-standard conditions have been recently published in Batz *et al.*[33] and are outside the scope of this paper. However the important MEOP features and results obtained at high magnetic field are reported here for clarity.

- Among the available OP atomic transitions for $^3$He at high magnetic field, the transitions f$^{2m}$ (see Abboud *et al.*[29] for notation) appear to give the best polarization values.

- For a given pressure above 10 mbar, the achievable polarization and its build-up time are increasing with the strength of applied magnetic field.

- The density of the $^3$He metastable state is not homogeneous inside the OP cell for pressure values above 32 mbar and a pump laser with an annular shape polarizes more rapidly, providing also higher polarization values[32].

- A dramatic increase in polarization value (up to 60 and 26.5 %) was obtained at 4.7 T in sealed cells filled with $^3$He at pressures of 67 and 267 mbar respectively.

Obtaining high polarization values at higher pressure is of great interest for building a compact polarizer, as the compression process is eased and the magnetization production rate is improved. To characterize the enhancement in magnetization production rate and assuming that the polarization builds-up exponentially with a time constant $t_b$, the following quantity R is introduced:

$$R = \frac{M_{eq} \times P}{t_b}, \tag{1}$$

where $M_{eq}$ is the equilibrium nuclear polarization and P is the pressure. The polarization build-up is found to deviate from a pure exponential and $t_b$ increases slightly when M is approaching its equilibrium value[33]. However, the deviation can be neglected in a first approximation. R is proportional to the average production rate of polarized atoms per unit volume over the time $t_b$ and is expressed in mbar/s. In the best standard conditions (few mbar and low magnetic field), the maximum R values are around 0.16 mbar/s when it was found to reach 0.5 mbar/s at 67 mbar and 1.5 T[30]. In other words, the same magnetic moment production rate could be obtained in a three times smaller volume at high magnetic field.

The present paper reports the first tests of a MEOP polarizer working inside a commercial MRI scanner at 1.5 T, whose construction was motivated by the results obtained for non-standard conditions as summarized above. The main expected benefit of having such a polarizer is to produce directly on-site the required amount of hyperpolarized $^3$He for lung MRI. As the magnetization production rate is higher at higher pressure, the duration of accumulation in the storage cell should be shorter. The compression factor required should be reduced and the corresponding polarization losses during the compression stage should also be lower. Moreover, no complicated setup to produce the guiding field is required, as it is already provided by the scanner. The magnetic field of 1.5 T was chosen due to the wide availability of scanners at this field strength, and because a good trade-off between high $M_{eq}$ and R values was achieved at this field. Firstly, the results of preliminary systematic studies of MEOP performed at 1.5 T with an annular laser beam shape and with different small sealed and open cells are presented. Their influence on the polarizer dimensions and design is discussed in section III. Finally, the results of first applications to lung MRI are presented.

## II. Preliminary study

### A. In sealed cells

Results obtained at 2 T with different shapes of laser beam demonstrated that MEOP could be made more efficient and reach higher polarization values when using an annular beam profile, created by a pair of conical lenses - axicons[32]. In order to determine the high-field polarizer's dimensions, expected flow rates and other characteristics, the MEOP systematic studies performed at 2 T with the implementation of the axicons[32] were repeated at 1.5 T. Six sealed cells of volume ~ 20 mL (1.5 cm diameter and 11 cm long) filled with pure $^3$He at different pressures (1.33, 32, 67, 96, 128, 267 mbar) were used for the experiments. Metastable atom densities of the order of $10^{10}$-$10^{11}$ atoms/cm$^3$ were created by a plasma discharge at 2 MHz. Experiments were carried out in a 1.5 T superconducting magnet (Magnex Scientific) with a 10 W (at a wavelength of 1083 nm) Keopsys Laser. The experimental arrangement, the optical detection of polarization's method and data analysis were similar to that used in the previously cited studies[31, 32]. Each cell was tested at a fixed pump power of 500 mW for 3 different plasma discharge conditions, and at a fixed density of the metastable state for 4 different laser power values (0.5, 1, 2 and 5 W), leading to six MEOP experiments per cell. The results of these 2 studies (influence of laser power and discharge condition) and additional features such as the variation of metastable atom density and $t_b$ value during the build-up process and the existence of a laser-induced relaxation are fully discussed in Collier[34].

From the six MEOP experiments performed at each pressure, the $M_{eq}$ and corresponding R values obtained during the experiment that was considered the most promising for building a high-field polarizer are reported in figure 1 (squares with solid lines). For comparison, previously published values obtained at 1.5 T with a Gaussian beam profile[30] (triangles) are shown. An increase of 25 % is observed in $M_{eq}$ values compared to the results obtained with a Gaussian beam, and a factor of ten improvement in the R values is obtained compared to the best standard conditions (low magnetic field and low pressure). As expected, the magnetization production rate increases with pressure, whereas achievable $M_{eq}$ decreases. The pair of axicons does not perform well at 1.33 mbar because the density of the metastable state is uniform inside the cell at this pressure. Hence, no improvements in the corresponding $M_{eq}$ and R were found compared to standard MEOP conditions. The most promising results seem to be obtained at 32 mbar, for which an R value of 1.417 mbar/s was reached and a polarization level of 66.4 % were achieved. The magnetization production rate can be increased at the expense of lower equilibrium nuclear polarization values, by using either higher laser powers, or increased densities of the metastable state. For a comparable gain in R value, the losses in achievable nuclear polarization were lower under the conditions of moderate laser power (0.5-1 W) and high density of metastable state ($n_m > 5 \times 10^{10}$ atoms/cm$^3$). That is why all the results shown in figure 1 were obtained under these conditions.

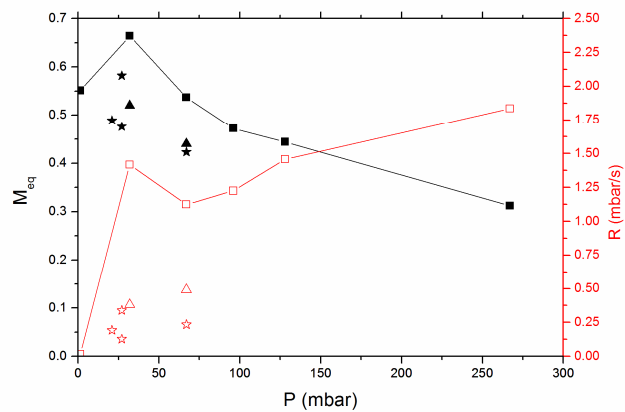

FIG. 1. Summary of the $M_{eq}$ (black filled symbols) and R values (red open symbols) obtained for different pressures at 1.5 T. Triangles and squares represent the values obtained in 20 mL sealed cells, with a Gaussian beam profile (published in Nikiel *et al.*[30]) and an annular beam shape respectively. The stars correspond to the prospective study performed in a 90 mL cell connected to a gas handling system, as discussed below.

**B. In a prospective open system**

To determine if the previous results could be reproduced in an open system, a dedicated gas handling system (GHS) was built to supply $^3$He to new small optical pumping cells located inside the 1.5 T superconducting magnet. The GHS consisted of a high purity bottle of $^3$He, followed by a getter filter, a 50 µm mechanical filter and a pressure meter to control the OP pressure. A bottle of $^4$He and a turbomolecular pump (TMP) were added in parallel before the pressure meter for cleaning purpose. The OP cell was connected to the GHS via a valve and a 4 m long electropolished non-magnetic stainless steel tube (6 mm outer diameter) to keep the TMP outside the fringe field of the magnet. The system was airtight and a vacuum of $10^{-7}$ mbar was maintained between experiments. The new OP cells were still 11 cm long but had different diameters of 16 and 31 mm (volumes of 20 and 90 mL, respectively). Each part of the system underwent a careful cleaning process. The cells, pipes and valves were heated up for few hours under the high vacuum. $^4$He was used to rinse the system and create high plasma discharge in the cells at low pressure, until the visible light emitted by the discharge contained only the wavelengths corresponding to the helium atomic transitions. Unfortunately and despite all these precautions, the purity of the gas could not be matched to that of sealed cells, which resulted in longer build-up time constants. Indeed, the presence of impurities increases the destruction rate of metastable atoms that are available for OP. Several MEOP experiments were performed in the 10-80 mbar range. The obtained $M_{eq}$ and R values for the 31 mm diameter cell are summarized in figure 1 (stars). Reasonable equilibrium nuclear polarization values of 40-60 % are presented. Although the corresponding R values are still higher than those for the standard MEOP conditions (0.1-0.4 mbar/s), the decrease in R compared to the experiments performed in sealed cells can be attributed to the lower $t_b$ values. This emphasizes the need to have a very high purity system, which would be more easily achieved with a non-magnetic GHS located as close as possible to the OP cell.

## C. Expected production rates

Assuming that the results ($M_{eq}$ and $t_b$ values) obtained during the preliminary studies can be reproduced in larger cells in a prototype high-field polarizer operating at 1.5 T, it is possible to derive the expected production rates for this system. In practice, the assumption can be matched if well designed capillaries are inserted at the input and output of the cell preventing backflow of helium and stopping impurities from diffusing from the output to the OP cell. It is also assumed that a steady state equilibrium has been reached and that the polarizer works with a constant flow Q. Considering the $^3$He diffusion coefficient in the 20-60 mbar pressure range, the dimensions of the cells and the average residency time of the gas in the cell, it can be assumed that the polarization is homogeneous inside the cell. Then, the magnetic moment produced in the OP cell per unit of time is equal to the quantity extracted by the compressor. Hence, it can be written:

$$\frac{dM}{dt} n = QM, \qquad (2)$$

where n is the number of atoms contained inside the OP cell expressed in standard cubic centimeter and Q is in sccm. It was found in Batz *et al.*[33] that the build up process of the nuclear polarization does not follow a pure exponential relationship under standard conditions. We observed the same phenomenom at 1.5 T[34]. If $t_b(M)$ represents the exponential time constant during the build up process, then $t_b(M)$ increases at the end of the experiment as M approaches the equilibrium value $M_{eq}$. The parameter $t_b$ used in equation (1) for the calculation of the R values refers to $t_b(0)$. As it is preferable to achieve high values of M in the high-field polarizer, the build up process will be assumed to adhere to the following equation:

$$M = M_{eq}\left(1 - \exp(-t/t_b')\right) \Rightarrow \frac{dM}{dt} = \frac{M_{eq}}{t_b'} \exp(-t/t_b') = \frac{M_{eq} - M}{t_b'}, \qquad (3)$$

where $t_b' = t_b(M_{eq})$. Empirically, $t_b'$ was found to be ~ 1.4 $t_b$[34]. By combining equations (2) and (3), the following equation (4) linking the chosen flow with the cell dimensions, build up time and resulting polarization can be obtained:

$$M = \frac{M_{eq}}{1 + \frac{Q \times t_b'}{n}}. \qquad (4)$$

In figure 2, the expected nuclear polarizations M as a function of the constant flow Q for a chosen OP cell volume of 1L have been extrapolated using equation (4) from the preliminary experiments' data in sealed and open cells.

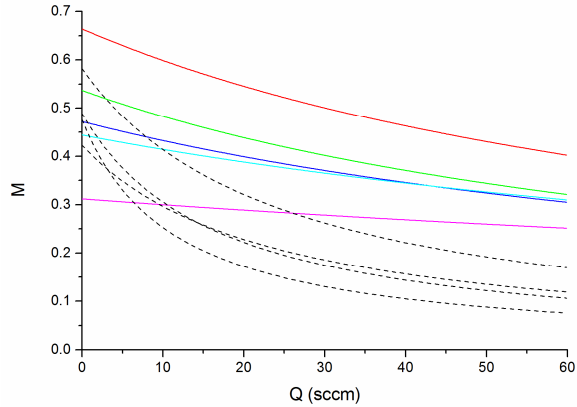

FIG. 2. Polarization values M, as a function of flow Q, derived from equation (4) with a chosen cell volume of 1 L. The colored solid lines represent an extrapolation of the results obtained in the preliminary studies in sealed cells at 32, 67, 96, 128 and 267 mbar. Each curve corresponds to a result displayed in figure 1 (black filled symbols) and can be identified by its corresponding M (Q = 0) = $M_{eq}$. In a similar manner, the black dashed lines represent an extrapolation of the results obtained in open cells with a GHS.

The expected nuclear polarization extrapolated from the open system experiments falls down much more rapidly with the increasing flow than that for the sealed cell experiments. It is due to the longer $t_b$' values; a consequence of the lower densities of metastable state in the presence of impurities. For higher Q, $^3$He atoms cannot be completely polarized during their average residency time in the OP cell. If the conditions of the 32 mbar sealed cell could be reproduced at a constant flow, figure 2 shows that only a 1 L OP cell would be needed to produce 60 sccm of $^3$He polarized at 40 %. In comparison, the large scale polarizer in Mainz needs a 36 L cell to produce the same flow of $^3$He polarized at 60 %. In a more realistic system similar to the one built for the prospective study, a 1 L of cell could produce 10-20 sccm of $^3$He with a nuclear polarization of approximately 30 to 45 %. The high-field polarizer reported in the present paper was designed to reach similar production rates and nuclear polarisations. Hence, the volume of the OP cells was chosen to be close to 1 L. Although better performance could be easily obtained by increasing the number or size of the OP cells, the prototype presented in the next section was developed as a proof of concept. For lung MRI, a reasonably good ventilation image requires an $^3$He dose of 90 sccfp (standard cubic centimeter fully polarized). It corresponds to a volume of 300 mL polarized to 30 %, which is a typical experimental protocol[35, 36]. For the target production rate of the high-field polarizer, this dose should be obtainable in less than 20 min.

## III. High-field polarizer design

The MEOP high-field compact polarizer is depicted in figure 3 and can be divided into four main parts. The main body of the polarizer is the optical pumping table that lies on the bed of the MRI scanner. It was designed to fit inside a birdcage coil dedicated to $^3$He human lung imaging, such that the coil could be used to measure the polarization inside the storage cell. It is compact and only two persons are required to carry the polarizer the short distance from the storage room to the scanner room. It is connected permanently using a flexible pipe to a small gas handling system located as close as possible to the OP cells, to avoid delicate cleaning of long tubing. At the end of an accumulation, the storage cell (mounted with the compressor on a separate surface) can be detached from the main body of the polarizer and kept inside the homogeneous magnetic field of the scanner, until the OP table is removed from the bed and a patient is ready to inhale a dose of hyperpolarized helium. Thus, the polarization losses due to magnetic field gradients are negligible. When the polarizer is not in use, it is connected to a cleaning system inside the storage room to avoid its contamination with impurities. The designs of each part of the polarizer are further discussed individually.

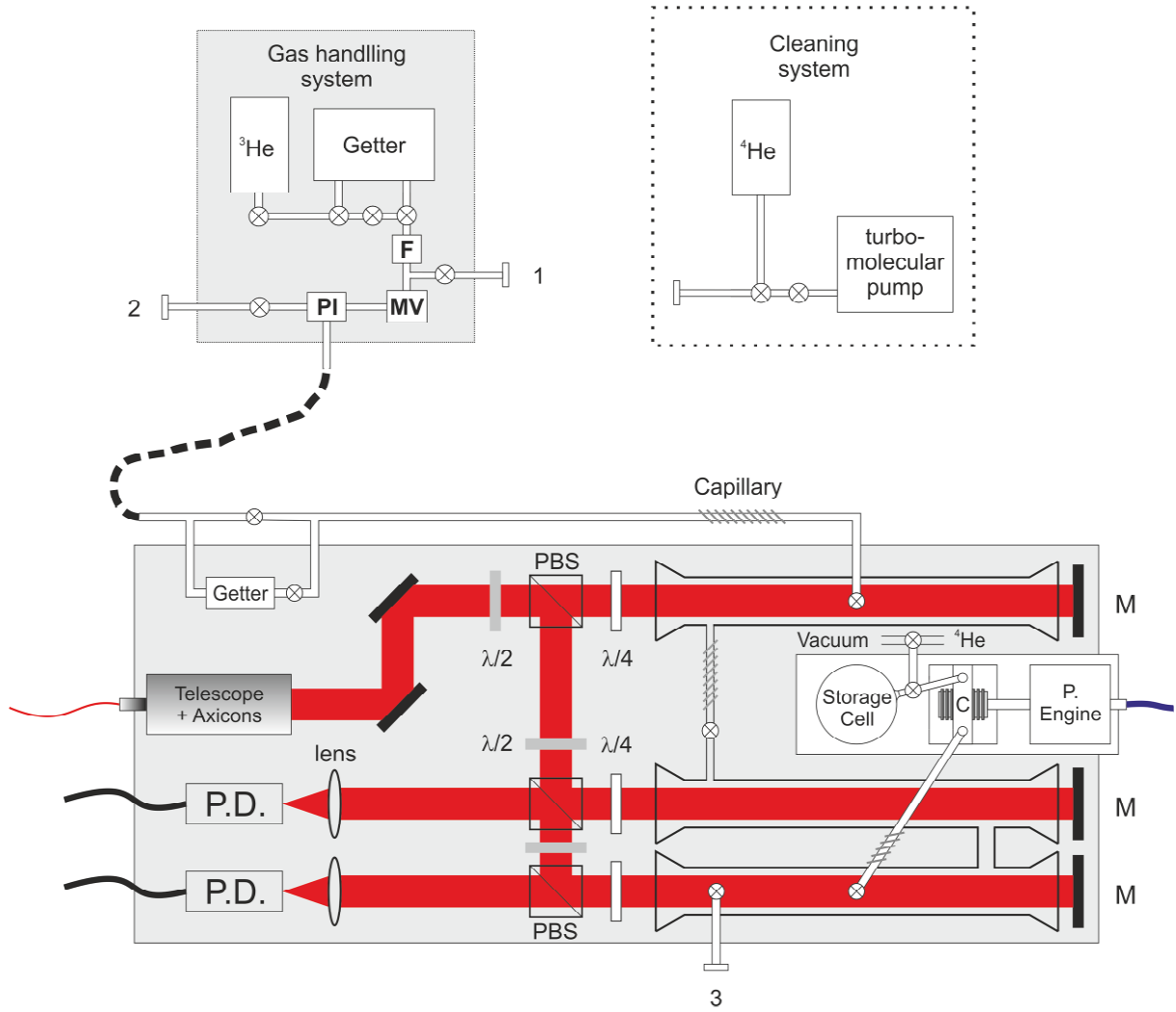

FIG. 3. Schematic of the high-field MEOP polarizer. Legend: see text.

## A. Optical pumping table

The wooden optical pumping table is 1.5 m long and 0.35 m wide. Three OP cells (internal diameter of 24 mm and length of 80 cm for a total volume of 1.1 L) are mounted in series. The diameter was chosen empirically as a compromise between a limited OP cell length required for a volume of ~ 1 L, and the difficulty to obtain an intense plasma discharge. Three capillaries are inserted for flow restrictions at the input of the first cell, between the first and the second cell and at the output of the third cell. The first capillary prevents any backflow of polarized helium to the GHS. The second one allows the total volume of the OP cells to be considered as being divided into two different compartments. When the polarizer is running with a constant flow, the polarization inside the first cell is different than in the two other cells (for which free diffusion induces a homogeneous polarization distribution). Consequently, two different regimes of OP (discharge intensities and laser powers) can be applied in the two compartments. The third capillary prevents impurities from diffusing from the peristaltic compressor to the OP cells. The choice of the capillary dimensions (length of 7 cm and diameter of 1.6 mm) results from a compromise between a low probability for $^3$He to diffuse in the opposite direction of the flow, and an acceptable pressure drop through the capillary. The calculations were inspired by a private communication from Pierre-Jean Nacher (Kastler Brossel Laboratory, ENS Paris) and are detailed in Collier[34].

The rest of the available space on the OP table is dedicated mainly to the optical elements. A 10 W laser from Keopsys (Lanion, France) operating at the 1083 nm wavelength (corresponding to $^3$He $2^3S_1$-$2^3P$ transition) is stored outside the fringe field of the magnet. The 4 m long laser fiber is connected to a Kepler-like telescope (magnification 2x) and a pair of axicons mounted on the OP table. The diameter of the resulting annular laser beam can be tuned from 18 to 28 mm to match the dimensions of the OP cells. The first pair of mirrors, placed after the axicons at 45° to the laser beam, regulates precisely the height and inclination of the beam. The beam is later is divided into three by a set of half-wave plates ($\lambda/2$) and polarization beam splitters (PBS). Each beam, after having been circularly polarized by a set of quarter-wave plates ($\lambda/4$) and aligned with each cell, passes through the cell back and forth by the means of additional mirrors (M). The transmittance of the second and third beams through their respective cells can be recorded by two photodiodes (P.D.). The transmittance value is used at the beginning of an experiment to tune the laser wavelength to the $f^{2m}$ transition.

The $^3$He plasma required to populate densities of the metastable state inside the OP cells, is created by high power radiofrequency (RF) discharges at 1 MHz. Two generators and two amplifiers are stored beyond the 5 gauss line of the magnet in order to obtain two different regimes of discharge in the first cell, and the other two. The high voltage is supplied to forty circular electrodes wound around each cell, with alternate polarity and 2 cm spacing between them. This configuration was found to be the most efficient to produce a dense plasma inside a 24 mm diameter cell. As previously discussed, it is possible to obtain a shorter build up process and a higher magnetization production rate by increasing the laser power and discharge intensity, at the expense of equilibrium polarization. The regime for the 1st OP cell was chosen to match these conditions, with a high RF discharge and a laser power of 3 W. The gas leaving the 1st cell can be considered as being "pre-polarized" before entering the 2nd and 3rd cells, where a second regime is sustained with 1.5 W of laser power per cell and moderate discharge. For this regime, the gas residency time is twice as long and higher polarization values are achieved, but with a longer time constant. Thus, having two different regimes results in a higher efficiency than if the same conditions were applied in all three cells.

**B. Storage cell and compression**

The peristaltic compressor (C), driven by a pneumatic engine (P. Engine), extracts the gas from the third cell into the 500 mL storage cell. These three elements are mounted separately on a separate surface (60 x 12 cm$^2$) that can be easily disconnected from the main optical pumping table when the compression is finished and the polarizer is removed from the scanner. When a subject is ready to be imaged, the hyperpolarized gas is transfered or compressed (by the same peristaltic compressor) from the storage cell into a Tedlar bag and mixed with $^4$He. A vacuum membrane pump and a bottle of $^4$He are also connected to the storage cell for rinsing between experiments. A pressure meter, not represented in figure 3, is used to measure the pressure inside the storage cell. The peristaltic compressor was designed by our group and is a smaller version of the one used in our low field polarizer[24]. It is mainly composed of polycarbonate, Plexiglas, polyamide and non-magnetic steel. Its dimensions were chosen to achieve a volumetric flow for the experimental range of pressures (20-40 mbar), matching the chosen values of Q (10-20 sccm) when running at 3-5 Hz. At this rotational speed, the vacuum level reached at the input of the compressor is on the order of $10^{-5}$ bar, for a 1 atm output pressure. To drive the peristaltic compressor, a non-magnetic pneumatic engine was purpose built by the Globe Airmotor BV company[37] to be used inside a high magnetic field. After succesful tests inside our magnet, the model was finalized and is now commercially available. The engine is driven by compressed air (2-3 bar) supplied by an air compressor.

**C. Gas handling system**

The GHS of the prospective study was modified and reduced in size to fit on a separate, compact Plexiglas surface of dimensions 60 x 50 cm$^2$. It consists of a high purity $^3$He bottle, a getter filter, a 50 μm filter (F), a fine metering valve (MV) to regulate the gas flow, a pressure meter (PI), and several valves. It is connected permanently to the main table via a 61 cm long flexible tube and lies on the bed of the scanner next to the main OP table during accumulation (figure 4). An additional compact getter was added and mounted on the OP table as close as possible to the entrance of the first cell for purity reasons. All the elements of the GHS and OP table are mainly non-magnetic and were checked to ensure normal operation at 1.5 T.

**D. Cleaning system**

A second part of the GHS, dedicated to the purpose of cleaning the system, was built separately. It consists of a TMP (producing a vacuum lower than $10^{-7}$ mbar) and a $^4$He bottle and is located in a storage room close to the MRI scanner. When the polarizer is not in use, it is connected to this cleaning system via a KF 16 flange adaptor. There are three locations where the cleaning system can be connected, labeled "1", "2" and "3" on figure 3. Location "1" was used mainly for cleaning the tubing of the GHS after it was built. Usually, the cleaning system is connected in parallel to "2" and "3". When it is required (for example after a long period of polarizer inactivity), the OP cells are filled with $^4$He at low pressure, and strong plasma discharges are created to remove impurities from the glass surface of the cells, before subsequently vacuuming the gas with the TMP. After the completion of the polarizer, it underwent a standard procedure of cleaning. The tubing and OP cells were heated to 100 ºC and vacuumed for few days. Then, the alternative high discharge / high vacuum procedure was repeated until the visible light emitted by the discharge contained only the wavelengths corresponding to the helium atomic transitions.

Pictures of the high-field polarizer working inside the clinical MRI scanner are displayed in figure 4. The polarizer is transported from the storage room to the scanner on a special non-magnetic tray.

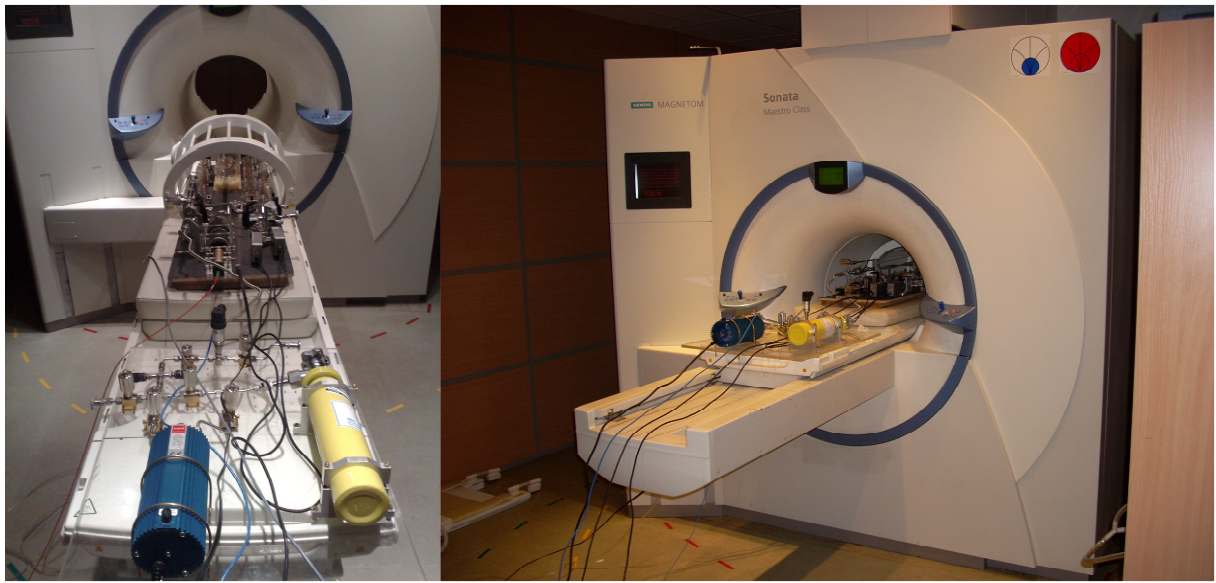

FIG. 4. Pictures of the high-field polarizer outside (left) and inside (right) the MRI scanner. The storage cell is located in the center of the birdcage coil and the GHS system is positioned at the bottom of the bed, which allows manual regulation of the gas flow during an accumulation experiment.

## IV. Results

After its completion, the polarizer was transported to the John Paul 2$^{nd}$ Hospital in Krakow where it could be tested inside a 1.5 T Sonata Siemens medical scanner. The scanner software was upgraded and a birdcage lung coil (Rapid Biomedical) was purchased, such that it was possible to run the system at the $^3$He resonance frequency (48.5 MHZ at 1.5 T).

**A. Characterization of the polarizer and first accumulations**

The method of optical detection of nuclear polarization[31] was unfortunately too complicated to implement on the polarizer in a clinical environment. Hence, M(t) and $M_{eq}$ could not be measured. However, the $t_b$ values could be estimated by recording the transmittance of the laser beam through each OP cell with a set of lenses and photodiodes. For a pressure of 20-30 mbar, the $t_b$ was found to be approximately 10-15 s in the first cell (with a high discharge and a 3 W laser beam) and about 25-30 s for the 2$^{nd}$ and 3$^{rd}$ cells (with 1.5 W of laser power and intermediate discharge intensity). These values were similar to those obtained previously for a sealed cell at 32 mbar (14-30 s) and lower than during the prospective study for open cells (45-100 s in the 20-30 mbar range).

A thermally-polarized phantom was purpose made for calibration of the polarization inside the storage cell. It consists of a 250 mL vessel, filled with 1.363 bar of $^3$He and 0.44 bar of $O_2$. The oxygen is used to shorten the longitudinal relaxation time, $T_1$, of the 14.02 mmol of thermally polarized $^3$He. A $T_1$ of 2.8 s was measured. The phantom was first used to obtain a flip angle calibration of the RF pulse of a spectroscopy sequence. A comparison between the free induction decay (FID) signals measured from the phantom and the polarizer's storage cell after an accumulation experiment was performed to deduce the nuclear polarization of $^3$He inside the storage cell. Figure 5 presents a calibration experiment that was conducted on the same day, with the exact same cell location inside the coils and the same flip angle. To increase the SNR of the signal from the phantom, the signal was averaged 60 times, with a repetition time of 25 s. The difference in size between the storage cell (10 cm diameter) and the phantom (8 cm diameter) was neglected, meaning that the filling factor was assumed to be the same. A Fourier Transformation of the free induction decay was performed and the area under the frequency peak was used to compare the signals.

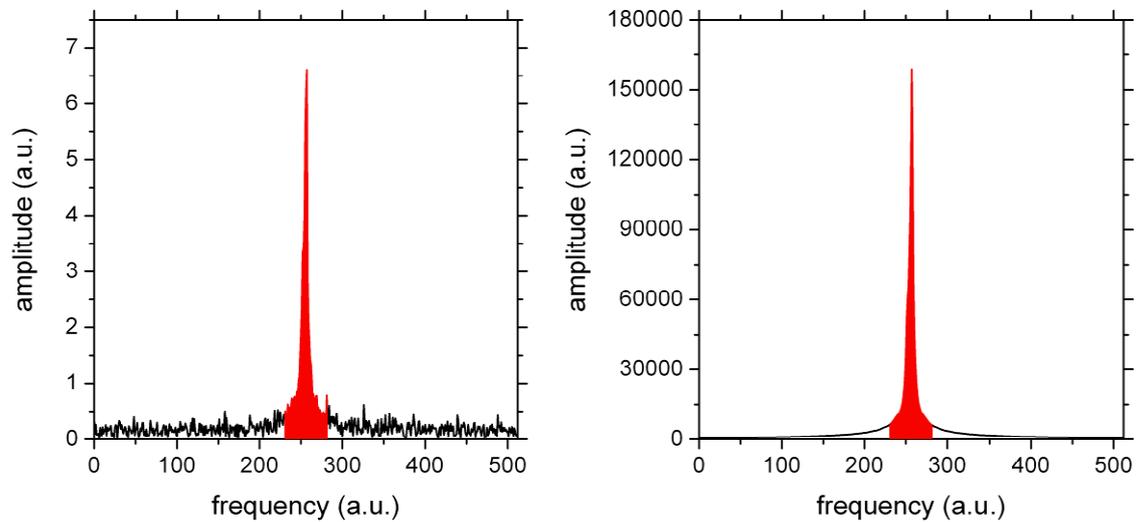

FIG. 5. Fourier Transform of the absolute value of the FID signal (512 samples, 10 kHz bandwidth) obtained from the thermally polarized phantom (left: theoritical M of $3.892 \; 10^{-6}$, n = 329.7 ± 7 scc, flip angle: 61.6 °) and the storage cell for the calibration experiment (right: n = 79,2 scc, flip angle: 61.6 °). A comparison of the integrated signal under the peak (area filled in red under the curve) yielded a polarization of 34.7 ± 1.5 % for $^3$He in the storage cell.

In total, nine accumulations were performed with n ranging from 29 scc for flow tests, to 493 scc for lung imaging experiments. As expected, it was found that increasing the flow reduced the polarization inside the storage cell. The typical output of the polarizer is a flow of 15 sccm with a polarization of 33 %. The highest M value (44.8 %) was obtained for a flow of 8 sccm. During some of the experiments, the relaxation time of the polarization inside the storage cell was measured to be $T_1$ = 208 ± 8 min, and a multinuclear multi-slice Spoiled Gradient Echo (FLASH) sequence was tested on the storage cell (see figure 6), demonstrating that lung MRI could be performed on healthy volunteers.

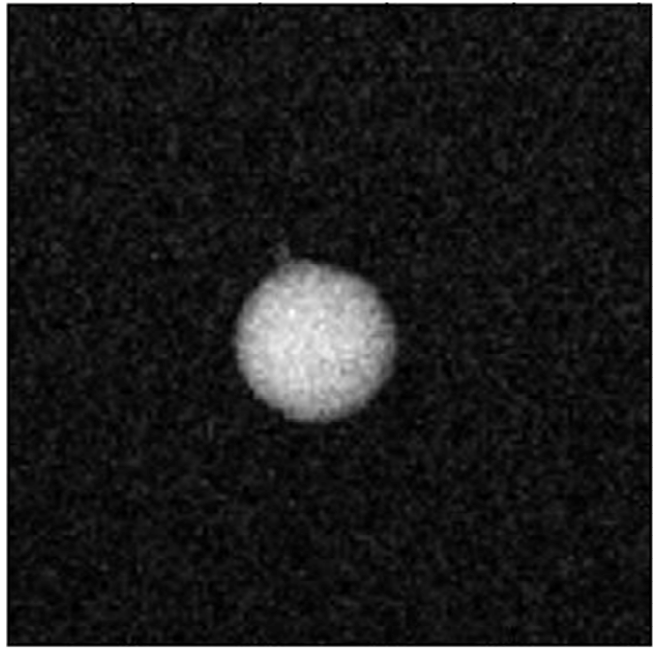

FIG. 6. Image of the 500 mL storage cell on the OP table and filled with 165.6 mbar of $^3$He polarized at 23.9 % (imaging parameters: resolution 128 x 128, FOV 400*400 mm$^2$, slice thickness 20 cm, TR 8 ms, flip angle 5.6°, acquisition time 1 s).

**B. Lung MRI**

Three accumulations of approximately 500 scc (corresponding to a pressure of 1 atm inside the storage cell) were performed with an average flow of 12.5 sccm (accumulation time of 40 min). After compression, the polarization was measured with a low RF flip angle. $^4$He was added until an absolute pressure of 2.4 bar was obtained and the storage cell was then closed. The peristaltic compressor was disconnected from the output of the 3$^{rd}$ OP cell and the optical table was removed from the scanner, while the storage cell was kept inside with the compressor and pneumatic engine. A healthy volunteer was introduced inside the scanner and the gas mixture being over the atmospheric pressure was released from the storage cell into a 1 L Tedlar bag (previously rinsed and pre-filled with $^4$He). The decay time $T_1$ inside the Tedlar bag was measured to be more than 1.5 h. After the sequence was prepared, the patient inhaled the mixture of $^3$He-$^4$He and the FLASH MRI sequence was performed during an apnea. The sequence lasted 1 s per slice for an image resolution of 128 x 128, limiting the apnea to only a few seconds. After the first image was obtained, the gas left inside the storage cell (almost half) could be extracted to the Tedlar bag via the peristaltic compressor for a second image to be taken. In figure 7, images acquired after two different accumulations are presented. Except a small asymmetry between right and left lung (presumably due to a hardware issue) the quality of the images is at the standard clinical level. The trachea and the first branches are clearly visible on these ventilation images of the lungs, for which promising SNR values of up to 65 were measured.

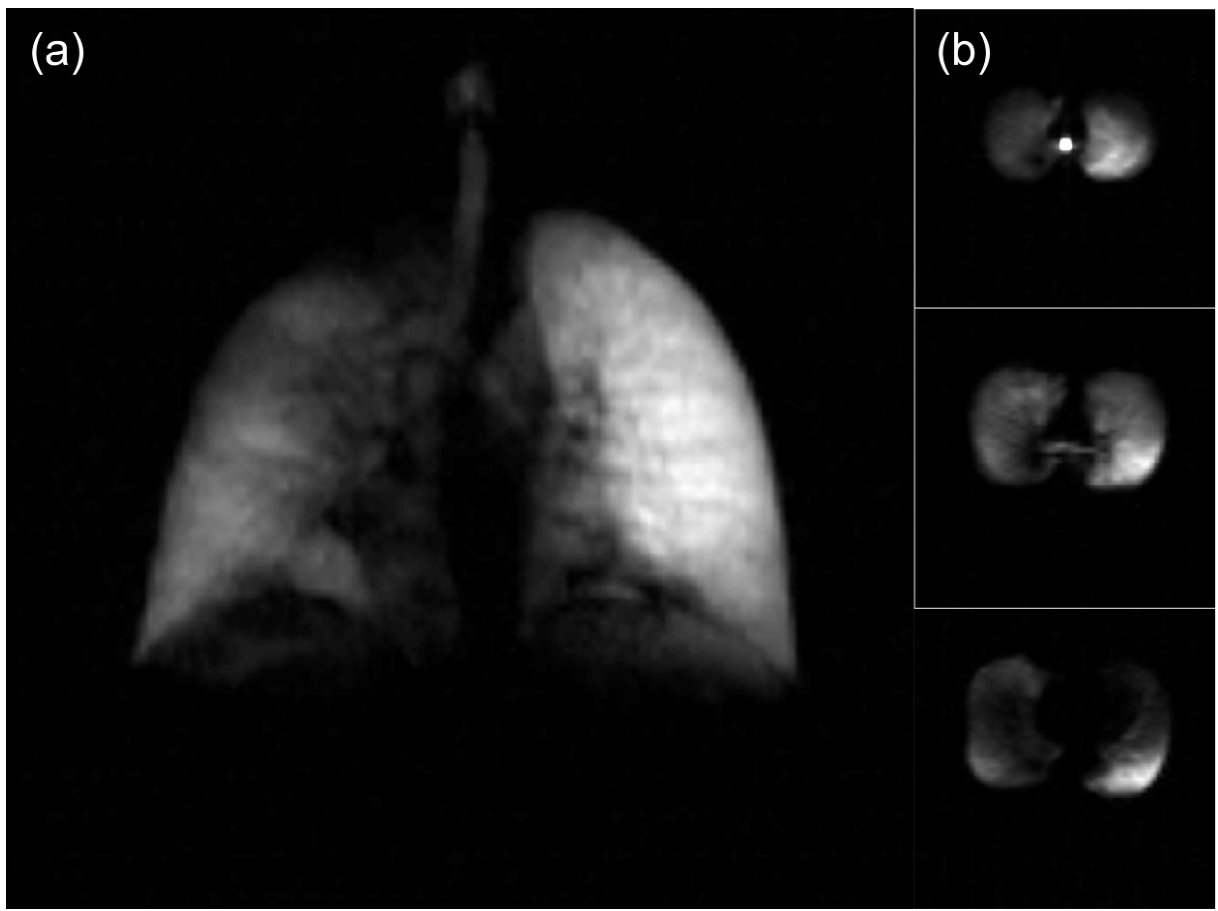

FIG. 7. (a) $^3$He (n = 272 scc with M = 20.8 ± 1 %) coronal image of the lungs of a healthy volunteer using a FLASH sequence (20 cm slice thickness, 38 cm FOV, 128 x 128 matrix, 8.6° flip angle, bandwidth per pixel 260 Hz, TE = 3.7 ms, TR = 7.9 ms, SNR = 56.3). (b) $^3$He (161 scc at 32.2 ± 1.5 %) transversal image the lungs of another healthy subject using a multi-slice FLASH sequence (5 cm slice thickness, 38 cm FOV, 64 x 64 matrix, 12.2° flip angle, bandwidth per pixel 260 Hz, TE = 3.7 ms, TR = 24 ms). Top to bottom: superior to inferior (SNR of 51.6, 67 and 44.1 respectively).

V. Conclusion

It has been demonstrated in this paper that a $^3$He high-field polarizer can be built and that MEOP can be performed at higher pressure and magnetic field than in standard conditions, in an open system. Although the polarizations and production rates are below those obtained in sealed cells (due to a difference in gas purity), the magnetization production rates reported are four times higher than those obtained under the best standard conditions. Furthermore, the high-field polarizer meets the requirements for providing a sufficient amount of hyperpolarized $^3$He for human lung medical studies. Typical gas flows of 10-20 sccm were obtained with corresponding nuclear polarizations of 30 to 45 %. These results agree relatively well with the expected production rates extrapolated from the open system prospective study and are, to our knowledge, higher than those of the other MEOP compact polarizer. Compared to other production systems, the gas is produced directly in-situ and there are negligible polarization losses due to storage and transportation of the gas. Another advantage of our system is its relatively low cost compared to polarizers with titanium alloy piston compressors.

However, the polarizer was designed as a prototype and could be further improved to make it more practical, with an automated system for controlling the valves and flow, a laser safety feature and a method for measurement of the polarization inside the different OP cells. It is also possible to increase the production rates further by increasing the number of OP cells, whose total volume is only 1.1 L in the present version. Further work will focus on each of these different aspects. However, the main inconvenience for a potential commercialization of such a high-field polarizer remains the global $^3$He shortage[38], responsible for a double issue of high price and limited availability.

**ACKNOWLEDGMENTS**

We gratefully acknowledge Pierre-Jean Nacher and Geneviève Tastevin from LKB in Paris for their support and advice. This work was partially supported by the Polish Ministry of Science and Higher Education (SPUB 547/6.PRUE/2008/07), the Marie Curie Research